\documentclass{aastex}
\usepackage{emulateapj5}

\shorttitle{Phase Separation in Giant Planets}
\shortauthors{Fortney and Hubbard}

\newcommand{\pf}{Paper I}
\newcommand{\mj}{$M_{\mathrm{J}}$}
\newcommand{\me}{$M_{\mathrm{E}}$}
\slugcomment{\bf}
\slugcomment{Accepted to ApJ, February, 2004}

\begin{document}

\title{Effects of Helium Phase Separation on the Evolution of Extrasolar Giant Planets}

\author{Jonathan J. Fortney\altaffilmark{1},  W. B. Hubbard\altaffilmark{1}} 

\altaffiltext{1}{Lunar and Planetary Laboratory, The University of 
Arizona, Tucson, AZ\ 85721-0092;
jfortney@lpl.arizona.edu, hubbard@lpl.arizona.edu}

\begin{abstract}

We build on recent new evolutionary models of Jupiter and Saturn and here extend our calculations to investigate the evolution of extrasolar giant planets of mass 0.15 to 3.0 \mj.  Our inhomogeneous thermal history models show that the possible phase separation of helium from liquid metallic hydrogen in the deep interiors of these planets can lead to luminosities $\sim$~2 times greater than have been predicted by homogeneous models.  For our chosen phase diagram this phase separation will begin to affect the planets' evolution at $\sim$~700 Myr for a 0.15 \mj\ object and $\sim$~10 Gyr for a 3.0 \mj\ object.  We show how phase separation affects the luminosity, effective temperature, radii, and atmospheric helium mass fraction as a function of age for planets of various masses, with and without heavy element cores, and with and without the effect of modest stellar irradiation.  This phase separation process will likely not affect giant planets within a few AU of their parent star, as these planets will cool to their equilibrium temperatures, determined by stellar heating, before the onset of phase separation.  We discuss the detectability of these objects and the likelihood that the energy provided by helium phase separation can change the timescales for formation and settling of ammonia clouds by several Gyr.  We discuss how correctly incorporating stellar irradiation into giant planet atmosphere and albedo modeling may lead to a consistent evolutionary history for Jupiter and Saturn.

\end{abstract}



\section{Introduction}
Over the past 8 years, nearly 120 giant planets have been found in orbit around other stars.  These planets have added immensely to the regions of parameter space in which we find giant planets and we are just beginning to understand how these interesting (and often hot!) environments affect the evolution of these objects. (See \citet{Hubbard02}, for a review.)  However, as we strive to understand these strange new worlds we must remind ourselves that our understanding of our closest two gas giants, Jupiter and Saturn, is far from complete.  As our examples of giant planets that will always be the most amenable for detailed study, it is of great importance to refine our understanding of these planets, and the physics that governs them, so that we will have confidence in our understanding of more distant giant planets.  As radial velocity studies reach longer time baselines, we are assured of finding planets similar to Jupiter and Saturn, and at similar orbital distances.  These wider orbital separation planets are also more likely to be directly imaged because they are farther from the glare of their parent stars.  This paper focuses on applying recent advances in our understanding of the evolution of Jupiter and Saturn to hypothetical extrasolar giant planets of various masses and orbital distances.

\begin{figure}
\plotone{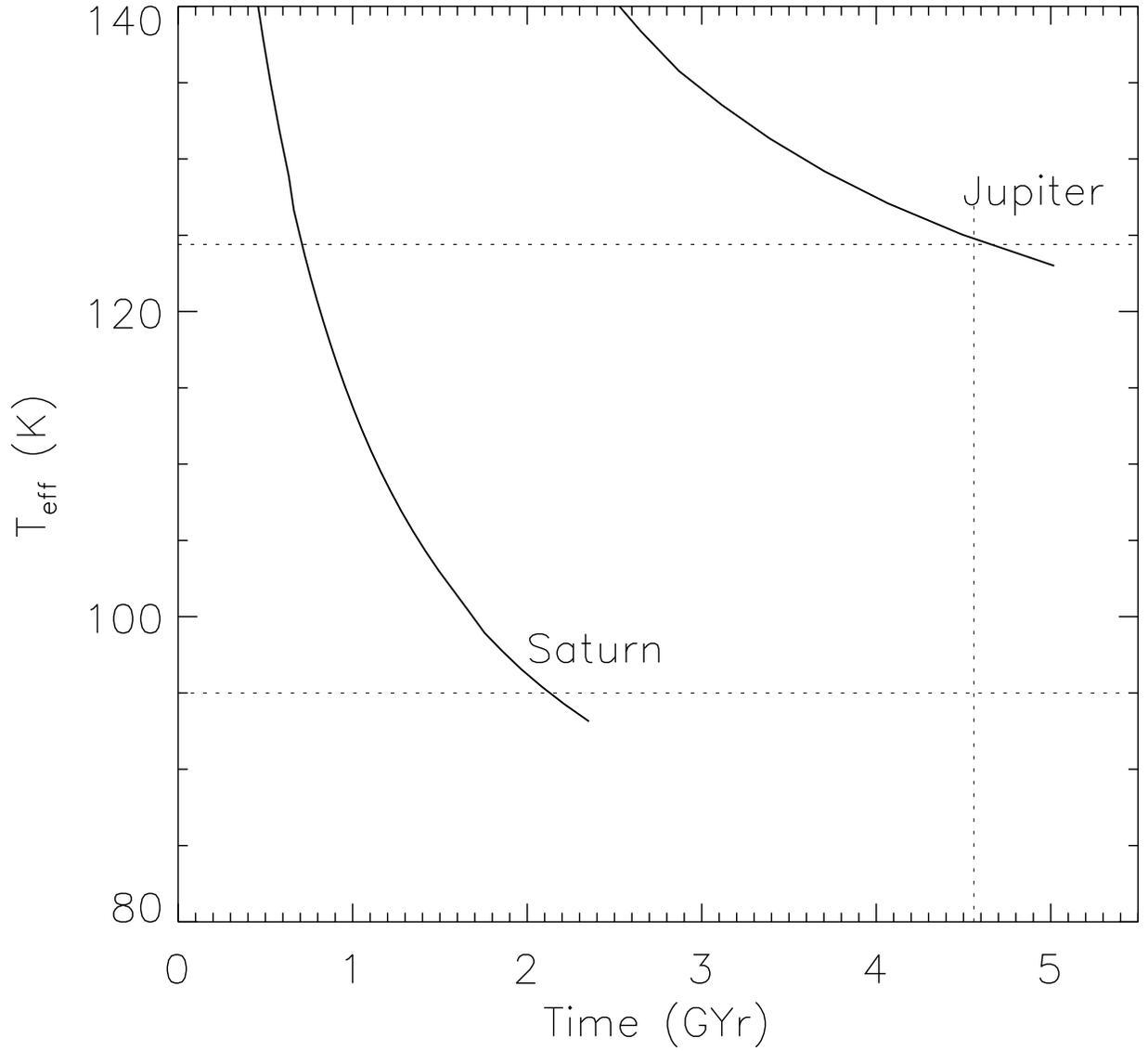}
\caption{Effective temperature vs. time in Gyr for homogeneous fully adiabatic models of Jupiter and Saturn from \pf.  Jupiter's core is 10 \me\ with a Z$_{envelope}$=0.059.  Saturn's core is 21 \me\ with a Z$_{envelope}$=0.030.  These values lead to the correct values of each planets' radius and moment of inertia at their known effective temperatures.  Both planets' known effective temperatures and the age of the solar system are shown as dotted lines.  Both planets possess $Y=0.27$.
\label{figure:js}}
\end{figure}

Our understanding of the evolution of Jupiter and Saturn is currently imperfect.  The most striking discrepancy between theory and reality is Saturn's luminosity. Saturn's current luminosity is over 50\% greater than one predicts using a homogeneous evolution model, with the internally isentropic planet radiating over time both its internal energy and thermalized solar radiation.  This discrepancy has long been noted \citep{Pollack77, Grossman80, Guillot95, Hubbard99}.  Homogeneous evolutionary models of Saturn tend to reach an effective temperature of 95.0 K (Saturn's current known $T_{eff}$) in only $\sim$~2.0 to 2.7 Gyr, depending on the hydrogen-helium equation of state (EOS) and atmosphere models used.  However, purely homogeneous models appear to work well for Jupiter.  \mbox{Figure~\ref{figure:js}} shows homogeneous evolutionary models for both planets from \citet{FH03} (hereafter \pf).  It has also long been believed that the most promising route to resolving this discrepancy is the possible phase separation of neutral helium from liquid metallic hydrogen in the planet's interior, beginning when Saturn's effective temperature reached $\sim$100 - 120 K \citep{SS77a,SS77b}.

Immiscibilities in two-component systems are common, and are the byproduct of the interaction potentials of the types of atoms (or molecules) in the mixture.  Once the temperature of a system becomes low enough, the energy of mixing becomes small enough that the Gibbs free energy of the system can be minimized if the system separates into two distinct phases.   One phase contains slightly less solute (here, helium) in the solvent (here, liquid metallic hydrogen) than there was initially, and the other phase nearly pure solute.  This is often termed immiscibility, insolubility, or phase separation, and the mixture is said to have a miscibility gap or solubility gap.  In general, the smaller the percentage of atoms in a mixture that is solute, the lower the temperature the mixture must attain for immiscibility to occur.  A common approximation \citep[see][]{Stevenson79, Pfaff} for the saturation value of $x$, the number fraction of the solute, is
\begin{equation} \label{x}
x=exp(B - A/k_bT),
\end{equation}
where $B$ is a dimensionless constant, $k_B$ is Boltzmann's constant, $T$ is temperature, and $A$ is a positive, pressure dependent constant with units of energy.  As described in \pf\ $B$ should be close to zero, and $A$ is the increase in free energy upon addition of a helium (or whatever atom in general) to pure liquid metallic hydrogen.  In \pf\ we showed that whether $A$ increases or decreases with pressure has important effects on giant planet evolutionary models.  In a solar composition mixture $x$ for helium is about 0.085.  Oxygen, the 2nd most abundant element, is down by a factor of over 100.  

\begin{figure}
\plotone{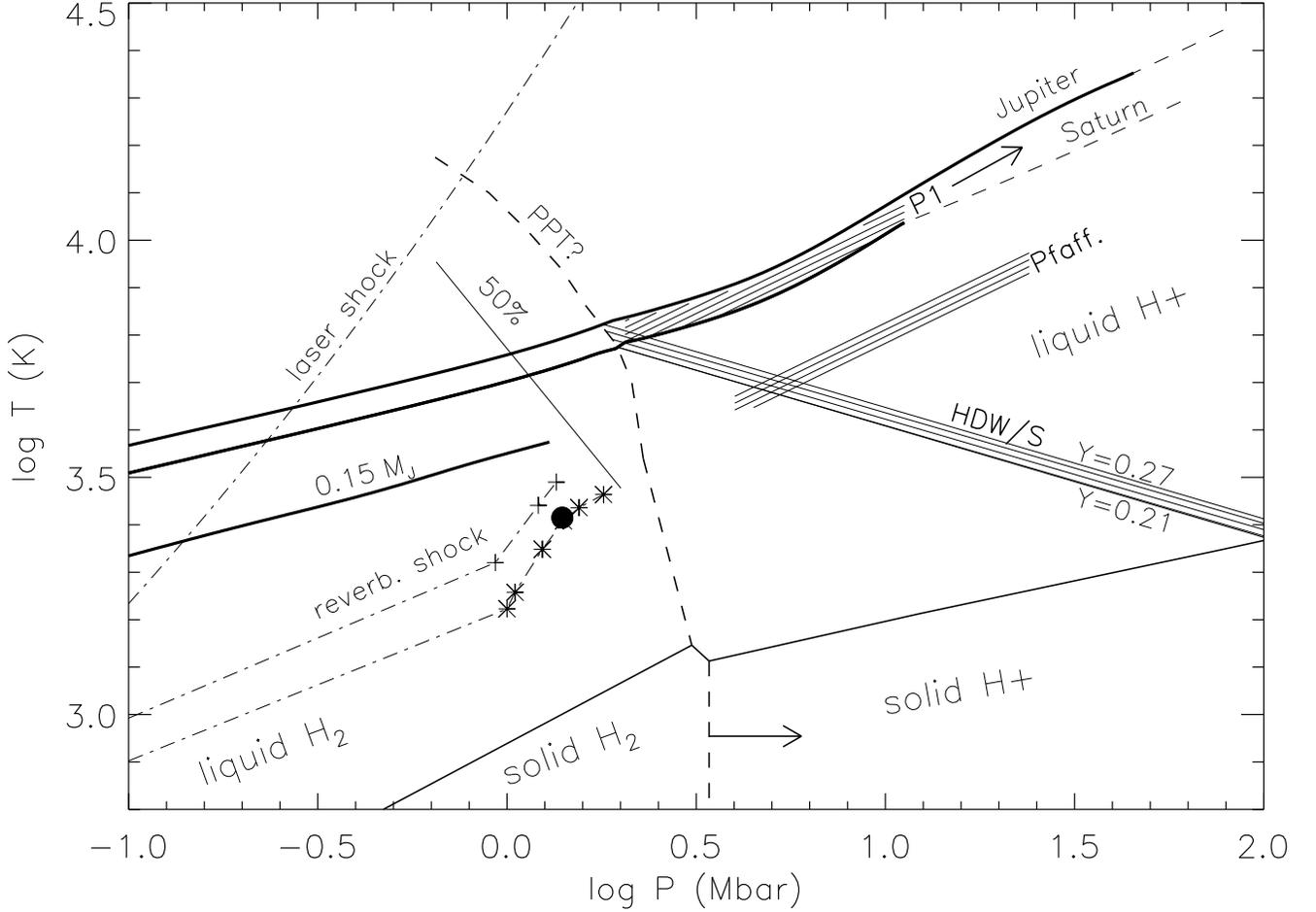}
\caption{Our current understanding of the high pressure phase diagram of hydrogen.  Regions of liquid molecular hydrogen (H$_2$) and liquid metallic hydrogen (H$^+$) are shown at high temperatures, and their solid counterparts at much lower temperatures.  The curve marked $PPT?$ is a possible transition from liquid H$_2$ to liquid H$^+$ as calculated by \citet{SCVH}. The solid line marked $50\%$ shows, for the theory of \citet{Ross98}, where liquid H$_2$ should be 50\% dissociated.  Laser shock data points from \citet{Collins98} are shown as a dash-dot line.  The reverberation shock data of \citet{Nellis99} are shown as pluses (for deuterium) and asterisks (for hydrogen).  The large black dot indicates the highest pressure that the conductivity of H$_2$ has been measured, which seems to indicate H$_2$ may be 10\% dissociated at this point \citep{Nellis99}. The calculated region of helium immiscibility from \citet{HDW} and \citet{Stevenson75} is labeled $HDW/S$.  The lines marked $Y=0.27$ and $Y=0.21$ mark the immiscibility boundaries for these two compositions.  The parallel lines labeled \emph{Pfaff.} show the helium immiscibility region as calculated by \citet{Pfaff}.  Again, the upper boundary is for $Y=0.27$ and the lower for 0.21, although they are not labeled to avoid clutter.  The current internal adiabats of Jupiter and Saturn are shown as heavy lines, while the dashed extensions show the pressure range within their cores.  Also shown in a heavy line is the adiabat of a hypothetical coreless 0.15 \mj\ planet at an age of 4.5 Gyr.  The parallel lines marked \emph{P1} between the Jupiter and Saturn adiabats is the ad-hoc immiscibility region from \pf.  The immiscibility lines are defined to be exactly parallel to the adiabats.  The arrow near \emph{P1} is meant to indicate that the immiscibility curves continue at this slope to higher pressures.
\label{figure:phased}}
\end{figure}

Since helium is relatively abundant in the hydrogen mixture, the helium (which is predicted to be neutral) will perturb the structure of the proton-electron plasma.  This $A$ constant has been calculated by various methods in the papers we will mention below to be $\sim$~1-2 eV, which for solar composition leads to temperatures on the order of 5000-10000 K for the onset of helium immiscibility.  (This is also dependent on pressure.)  \mbox{Figure~\ref{figure:phased}} shows in detail our current knowledge of the high pressure phase diagram of hydrogen and helium \citep{Hubbard02}.  Labeled are the current interior adiabats of Jupiter, Saturn, and a hypothetical 0.15 \mj\ planet.  Relevant experimental and theoretical boundaries are also labeled, as are regions of calculated helium immiscibility.

\citet{Salpeter73} was the first to note the effects of the immiscibility of helium in liquid metallic hydrogen on the structure and evolution of a hot, adiabatic, hydrogen-helium planet.  \citet{Stevenson75}, using perturbation theory, performed the first detailed calculation of the hydrogen-helium phase diagram, in an effort to map the regions of pressure-temperature-composition space in which helium was likely to become immiscible.  His calculations roughly agreed with estimates of the current pressures and temperatures of liquid metallic hydrogen in Jupiter and Saturn's interiors.  These calculations indicated that as pressure increased, the saturation concentration of helium in liquid metallic hydrogen would increase, leading to constant composition curves that slant down and to the right in \mbox{Figure~\ref{figure:phased}}.

Soon after, \citet{SS77a, SS77b} performed detailed calculations on the dynamics and distribution of helium in giant planets.  They found that when helium becomes immiscible in liquid metallic hydrogen, the composition that separates out is essentially pure helium, and this helium on fairly short timescales (relative to the convective timescale) will coalesce to form helium droplets.  These droplets, once they reach a size of $\sim$~1 cm, will attain a Stokes velocity greater than the convective velocity and will then fall down through the planet's gravitational field.  If the droplets reach a region where helium is again miscible at higher concentration, they will redissolve, enriching the deeper regions of the planet in helium.  They found that this ``helium rain'' could be a substantial additional energy source for giant planets.  Helium would be lost from $all$ regions with pressures lower than the pressures in the immiscibility region, since the planet is fully convective (or nearly so) up to the visible atmosphere.  Excess helium would be mixed down to the immiscibility region and be lost to deeper layers.  This would leave all molecular regions up to the visible atmosphere depleted in helium.

Relatively few studies have been done since then on phase diagrams of hydrogen-helium mixtures.  \citet{HDW}, using a Monte Carlo technique, but similar assumptions to that of \citet{Stevenson75}, obtained essentially the same results.  The most recent calculations \citep{Pfaff} utilized molecular dynamics to predict a helium-immiscibility region with a shape very different from that of \citet{Stevenson75} and \citet{HDW}.  They find that as pressure increases, the saturation concentration of helium in liquid metallic hydrogen decreases, which leads to constant composition lines that slant up and to the right in \mbox{Figure~\ref{figure:phased}}.  Interestingly, these constant-composition lines run nearly parallel to the giant planet adiabats.  Detailed inhomogeneous evolutionary models including helium phase separation were not performed until \citet{Hubbard99}.  These authors investigated the cooling of Jupiter and Saturn when the mass of helium rained out linearly with time since Jupiter and Saturn's formation, or alternatively, rained out just before the planets reached their known effective temperatures.  These are two logical limiting cases.

There is observational evidence that helium phase separation has begun in both Jupiter and Saturn.  The protosolar mass fraction of helium is calculated to be near $Y$=0.27 (\citet{Lodders03} puts the number at 0.2741).  The atmospheres of both Jupiter and Saturn are depleted relative to this value.  With the assumption that these planets globally contain the protosolar $Y$, the missing helium must be in deeper layers of the planet.  The case for Jupiter's depletion is clear cut, with a value of $Y=0.234 \pm .005$ from the Helium Abundance Detector (HAD) on the Galileo Entry Probe \citep{vonzahn98}.   The case for Saturn is much less clear.  Without a past or planned Saturn entry probe, Saturn's atmospheric helium abundance can only be obtained through indirect methods, using infrared spectra with or without radio occultation derived temperature-pressure \emph{(T-P)} profiles.  Analysis of Voyager measurements indicated $Y=0.06 \pm 0.05$ in Saturn \citep{Conrath84}.  However, the mismatch between the Voyager derived value for Jupiter ($Y=0.18 \pm 0.04$, \citep{Gautier81}) and the accurate HAD measurements, along with \citet{Hubbard99} evolutionary and \citet{Guillot99} static models, led \citet{CG00} to perform a reanalysis of the Voyager data.  The details of their investigation will not be described here, but by disregarding the occultation derived \emph{T-P} profile, which may be in error, they obtain $Y=0.18-0.25$ for Saturn's atmosphere.

\begin{figure}
\plotone{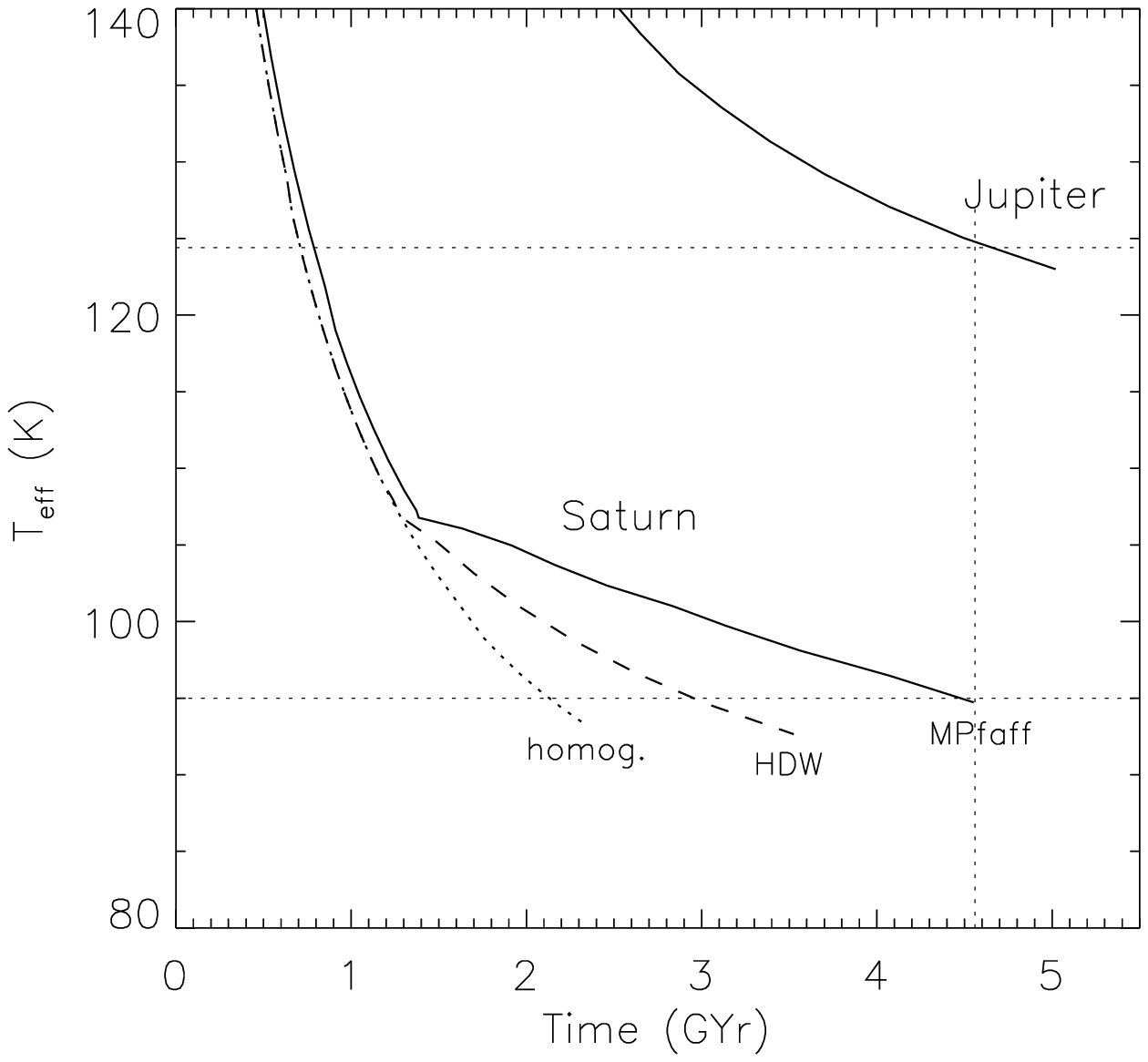}
\caption{Additional evolutionary models from \pf\ including the phase separation of helium from liquid metallic hydrogen.  The dotted curve for Saturn and the solid curve for Jupiter are the evolutionary tracks shown in \mbox{Figure~\ref{figure:js}}.  The dashed curve includes the phase diagram of \citet{HDW}, while the solid curve for Saturn uses the proposed ad-hoc phase diagram from \pf.  The Saturn curves are slightly offset at younger ages due to differences in initial core masses and heavy element abundances needed for the each planet to reach Saturn's known radius and moment of inertia at 95 K. The \pf\ phase diagram allows both Jupiter and Saturn to reach their known ages and $T_{eff}$s.
\label{figure:js2}}
\end{figure}

Noting the clear need to better understand Jupiter and Saturn in light of these atmospheric $Y$ values, in \pf\ we calculated the first evolutionary models that coupled high-pressure phase diagrams of hydrogen-helium mixtures and a grid of radiative atmosphere models for giant planets.  A variety of Saturn evolutionary models were calculated that included helium phase separation.  The main findings of \pf\ were as follows.  The phase diagram of \citet{HDW}, which is essentially the same as that of \citet{Stevenson75}, is inapplicable to the interiors of Jupiter and Saturn, if helium phase separation is Saturn's only additional energy source.  These phase diagrams predict that $A$ from equation (\ref{x}) is a decreasing function of pressure.  As \mbox{Figure~\ref{figure:js2}} shows, this phase diagram prolongs Saturn's cooling 0.8 Gyr, even in the most favorable circumstance that all energy liberated is available to be radiated, and does not instead go into heating the planet's deep interior.  \pf\ found that if one were to match the $Y_{atmos}$ of \citet{CG00} and prolong Saturn's evolution to a $T_{eff}$ of 95.0 K at 4.56 Gyr, the helium that becomes immiscible and rains down to deeper layers needs to rain far down into the planet, likely all the way to the core, in order for enough energy to be released and still match the relatively high ($Y_{atmos}$=0.18 - 0.25) helium abundance.

In \pf\ an ad-hoc phase diagram was created that was essentially a modification of the phase diagram calculated by \citet{Pfaff}.  (\citet{Pfaff} find a phase diagram in which $A$ from equation (\ref{x}) is an increasing function of pressure.)  To simplify the evolution of the planets the ad-hoc phase diagram was constructed such that helium immiscibility region runs exactly parallel to the planets' internal adiabats.  Therefore, there is no region for the helium droplets to redissolve in the liquid metallic hydrogen.  This causes helium that phase separates to rain all the way down to the planet's heavy element core.  Consequently, all hydrogen/helium regions, molecular and metallic, become more helium poor as the helium layer on top of the core grows.  This ad-hoc phase diagram allows Saturn to reach an age of 4.56 Gyr and $T_{eff}$ 95 K while its $Y_{atmos}$ drops to 0.185.  \mbox{Figure~\ref{figure:js2}} shows the evolution of $T_{eff}$ vs.~time.  With this phase diagram, Jupiter evolves homogeneously to the present day and reaches $\sim$~4.7 Gyr at 124.4 K without helium becoming immiscible.  Jupiter would then begin to evolve inhomogeneously at $T_{eff}$ below 123 K.  Still in need of explanation is Jupiter's depletion of helium (and neon, which may have been carried away in the helium \citep{Roulston95}) in its atmosphere.

What we have from \pf\ is a high-pressure hydrogen-helium phase diagram that is calibrated to Jupiter and Saturn.  Specifically, both planets reach their known effective temperatures after $\sim$~4.6 Gyr, with an improved but still imperfect understanding of helium phase separation.  The purpose of the present paper is to investigate the effects of helium phase separation on the evolution of theoretical giant planets in orbits around other stars using the phase diagram derived in \pf.

Here we concentrate on planets ranging in mass from 0.15 \mj\ (half Saturn's mass) to 3.0 \mj, and derive the luminosity, $T_{eff}$, radius, and $Y_{atmos}$ as a function of time during the planets' evolution.  Of the many extrasolar planets found to date, the planet with the smallest minimum mass is HD 49674b, at 0.12 \mj\ \citep{Butler03}.  As we will discuss later, a planet with a mass this small likely contains no liquid metallic hydrogen, only dense molecular hydrogen.  Approximately 70\% of all known planetary candidates have minimum masses of less than 3.0 \mj, so planets of the masses we explore here are sure to be abundant.

Our standard models (to be discussed in Section \ref{isolate}) incorporate a primordial 10 \me\ heavy element core for all planets, but later in Section \ref{cores} we investigate the effects of 20 \me\ and coreless models.  (1 Jupiter mass, \mj, is 317.7 Earth masses, \me.)  We will later find that for the more massive planets, varying the core mass has little effect on $T_{eff}$ but a large effect on planetary radii.  In Section \ref{irrad} we investigate the effects of modest stellar irradiation.  Stellar heating retards a planet's cooling, and we find that if a planet is within a few AU of its parent star, the planet will reach its equilibrium temperature before its interior reaches temperatures cool enough for helium to become immiscible.  We calculate the cooling of planets in isolation and at 10 and 5 AU from a constant luminosity 1.0 L$_{\odot}$ star.  In Section \ref{discuss} we discuss the atmospheric properties and detectability of these EGPs likely to be undergoing helium phase separation.

\section{Description of the Models} \label{models}
Our evolutionary models are the same as those described in \pf\ and use the method first described by \citet{Hubbard77}.  A thermal history calculation for an isolated nonrotating giant planet of mass \emph{M}, radius \emph{a}, and specified composition \emph{X} (where \emph{X} is a matrix of the various mass fractions of elements), yields relations of the form
\begin{equation} \label{e1}
L \equiv 4\pi\sigma a^2 T_{eff}^4 = L(M,t,X),
\end{equation}
\begin{equation} \label{a1}
a=a(M,t,X),
\end{equation}
where \emph{L} is the planet's luminosity, $\sigma$ is the Stefan-Boltzmann constant, $T_{eff}$ is the planet's effective temperature, and $t$ is the planet's age (i.e., time since accretion of its hydrogen envelope).  Under the assumption of homogeneous evolution, i.e., that \emph{X(r)} = constant, (where \emph{r} is the radius of a mass shell inside the planet) and therefore \emph{S(r)} = constant (where \emph{S} is the specific entropy of the deep interior), expressions (\ref{e1}) and (\ref{a1}) can be derived with the help of a grid of model atmospheres.  The grid is obtained by choosing independent variables $T_{eff}$ and \emph{g} (the atmosphere's surface gravity), integrating the atmospheric structure inward to a depth where it is fully convective and essentially isentropic, and then calculating \emph{S} of the atmosphere at depth:
\begin{equation}
\label{S}
S=S(T_{eff},g,X),
\end{equation}
where the surface gravity is given by
\begin{equation}
\label{g}
g=G M/a^2.
\end{equation}

For the hydrogen-helium envelope of the planet we use the ``interpolated'' equation of state (EOS) of \citet{SCVH}.  Our initial helium mass fraction is $Y=0.27$.  For the heavy element core we use the ANEOS olivine EOS.  For the heavy elements in the envelope (we take $Z=0.02$) we use the ANEOS water EOS.  We use the model atmosphere grid of \citet{Burrows97}, as updated for \citet{Hubbard99}.  It is a non-gray grid of atmospheres at solar composition, suitable for isolated giant planets and brown dwarfs.  We couple our phase diagram from \pf\ to our evolution models in the following way.  Using our phase diagram and a constant reference pressure of 2 Mbar, we calculate the specific entropies of adiabats for many $Y$ values, from 0.27 to 0.055, in steps of 0.005.  Our helium poor adiabats then have homogeneous helium-poor composition from the outer boundary (1 bar) until the beginning of the pure helium region.  The number of mass shells needed for the pure helium region is calculated under the constraint that helium must be conserved to 0.05\% or better for the entire planet.  The entropy of the pure helium region is set by the constraint that there be no discontinuity in temperature across the boundary.  The mass of the heavy element core remains the same for all models.

We make several approximations, most of which will affect our results to fairly  minor degrees.  The first is that the planets are spherically symmetric and nonrotating.  This is entirely justified as these planets are hypothetical and the rotation rates of non-tidally locked EGPs are entirely unknown.  For Jupiter and Saturn models, neglecting rotation introduces errors in evolution timescales of only a few percent.  We also assume that the primordial heavy element cores do not take part in the planets' evolution.  The vast majority of a giant planet's thermal reservoir is in the liquid metallic hydrogen, which has a high heat capacity.  The energy generated by radioactive decay in the core contributes negligibly to the heat budget of a giant planet \citep{Hubbard80}.

Our treatment of irradiation from the parent star is only approximate.  Recently, a move has been made to consistently incorporate the effects of stellar irradiation on the evolution of giant planets.  This has been shown to be of critical importance for close-in giant planets such as 51 Peg b and HD 209458b \citep[see][]{Guillot02,Baraffe03,Burrows03}.  For the planets we study here, which are at least 5 AU from their parent star, we incorporate the effects of irradiation using an approximation that has commonly been used for Jupiter and Saturn.  We assume the incoming stellar radiation that is scattered does not change the planetary atmosphere's \emph{T-P} profile from that of an isolated object.  Furthermore, the stellar radiation that is absorbed (thermalized) is absorbed down in the convective region of the planet's atmosphere, which shares the adiabat of the deep interior.  In this way we can use a grid of model atmospheres for isolated objects.  Following \citet{Hubbard77}, equation~(\ref{e1}) must be modified as below.
\begin{equation}
\label{Lmod}
L \equiv 4\pi\sigma a^2 (T_{eff}^4 - T_{eq}^4) = L(M,t,X),
\end{equation}
where $T_{eq}$ is the effective temperature that the planet would have if it had no intrinsic luminosity 
($L=0$).  We derive $T_{eq}$ from the Bond albedo, $A_B$, according to
\begin{equation}
\label{Teq}
4\pi\sigma a^2 T_{eq}^4 = (1-A_B)\pi a^2 L_{\star}/4 \pi d^2,
\end{equation}
where $L_{\star}$ is the stellar luminosity and \emph{d} is the star-planet distance.  Using equation~(\ref{Lmod}), an equation can be written that gives the heat extracted from the planet's interior per unit time:
\begin{equation}
\label{LS}
L(M,t,X)=-M\int_{0}^{1}dm T \frac{\partial S}{\partial t},
\end{equation}
where the dimensionless mass shell variable \emph{m} is defined by
\begin{equation}
\label{shell}
m=\frac{1}{M}\int_{0}^{r} 4 \pi r^{{\prime}^2} dr^{\prime} \rho(r^{\prime}).
\end{equation}
Put simply, a planet's luminosity is derived directly from the decrease in entropy of its interior.  Equation~(\ref{LS}) is valid for either homogeneous or inhomogeneous evolution, because \emph{S} is a function of the mass shell.  (For homogeneous evolution, since \emph{X} does not vary with the mass shell, neither does \emph{S}.) Regions of the interior that are helium rich have a lower entropy per unit mass.  As the planet cools and the helium region on top of the core grows, more mess shells are ``converted'' from hydrogen-helium mixtures (high entropy) to pure helium (low entropy).  We can then calculate the time step between successive models in an evolutionary sequence.  Equation (\ref{LS}) can be rewritten as
\begin{equation} \label{time}
\partial t = - \frac{M}{L} \int_{0}^{1}T \partial S dm,
\end{equation}
where $\partial t$ is the time step, and the other variables have the same meanings as defined earlier.

In our calculations we assume that there is no discontinuity in temperature across the boundary between the helium depleted and pure helium regions.  From the \citet{SCVH} EOS we calculate the adiabatic temperature gradient in the pure helium region and ensure that the temperatures of the two regions match at the composition boundary.  In general this prescription may be too simple.  As discussed in detail in \citet{SS77b} and also in \pf, a helium composition gradient will lead to an increased temperature gradient in order to maintain convective instability.  Here we assume an infinitely thin boundary between our two regions, but that is an artifact of the shape of our phase diagram.  More generally, for arbitrary phase diagrams, helium composition gradients would occur with effects on the temperature gradient.

\section{Results of Calculations} \label{results}
\subsection{Evolution of Isolated EGPs} \label{isolate}
\begin{figure}
\plotone{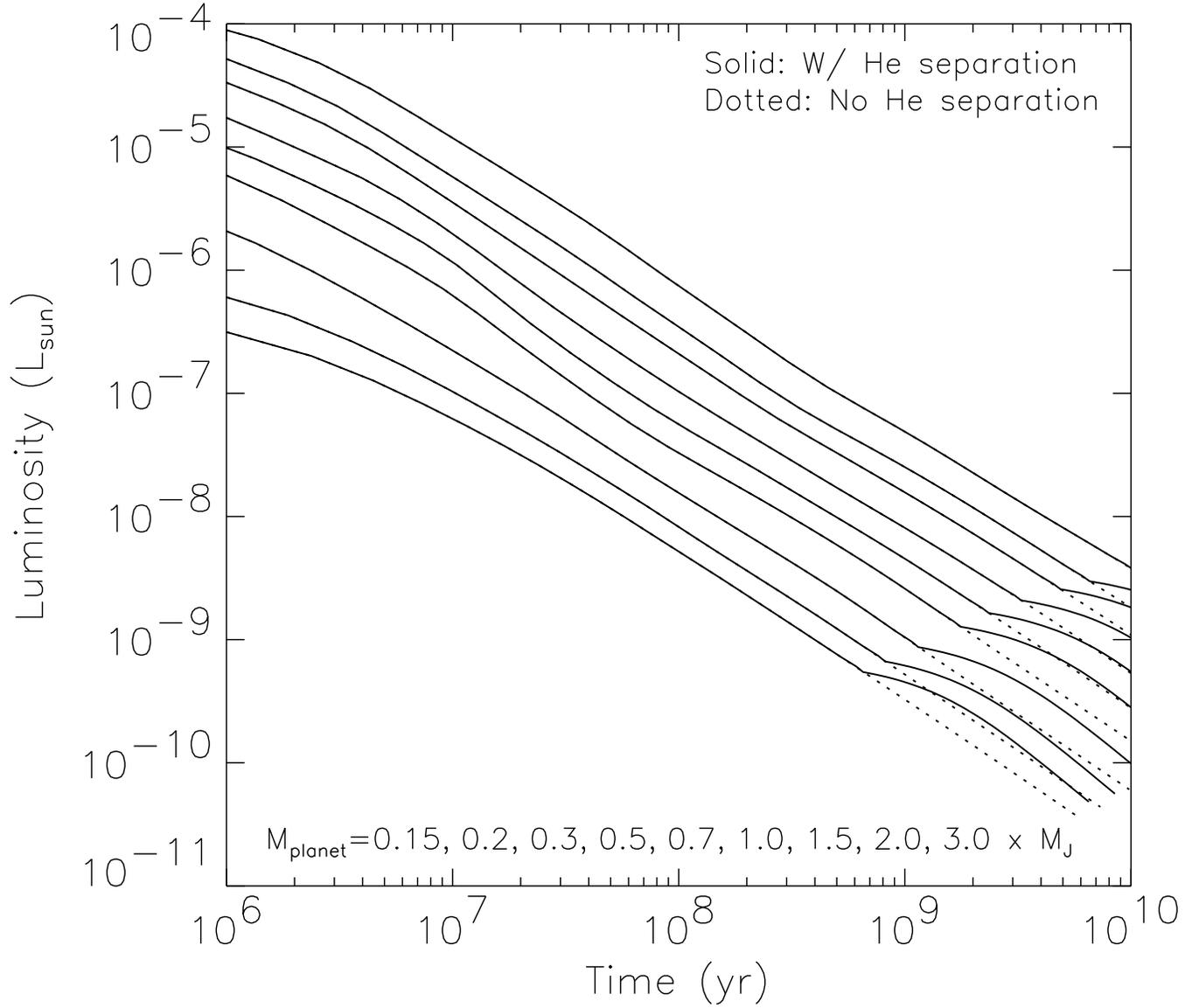}
\caption{Evolution of the luminosity for planets of mass 0.15 to 3.0 \mj\ for our standard models with no stellar irradiation and 10 \me\ cores.  The dotted lines are models without helium phase separation, while the solid lines include the effects of helium phase separation on the planets' cooling.  The top curve is for the highest mass planet while the bottom curve is for the lowest mass planet.
\label{figure:L}}
\end{figure}

\begin{figure}
\plotone{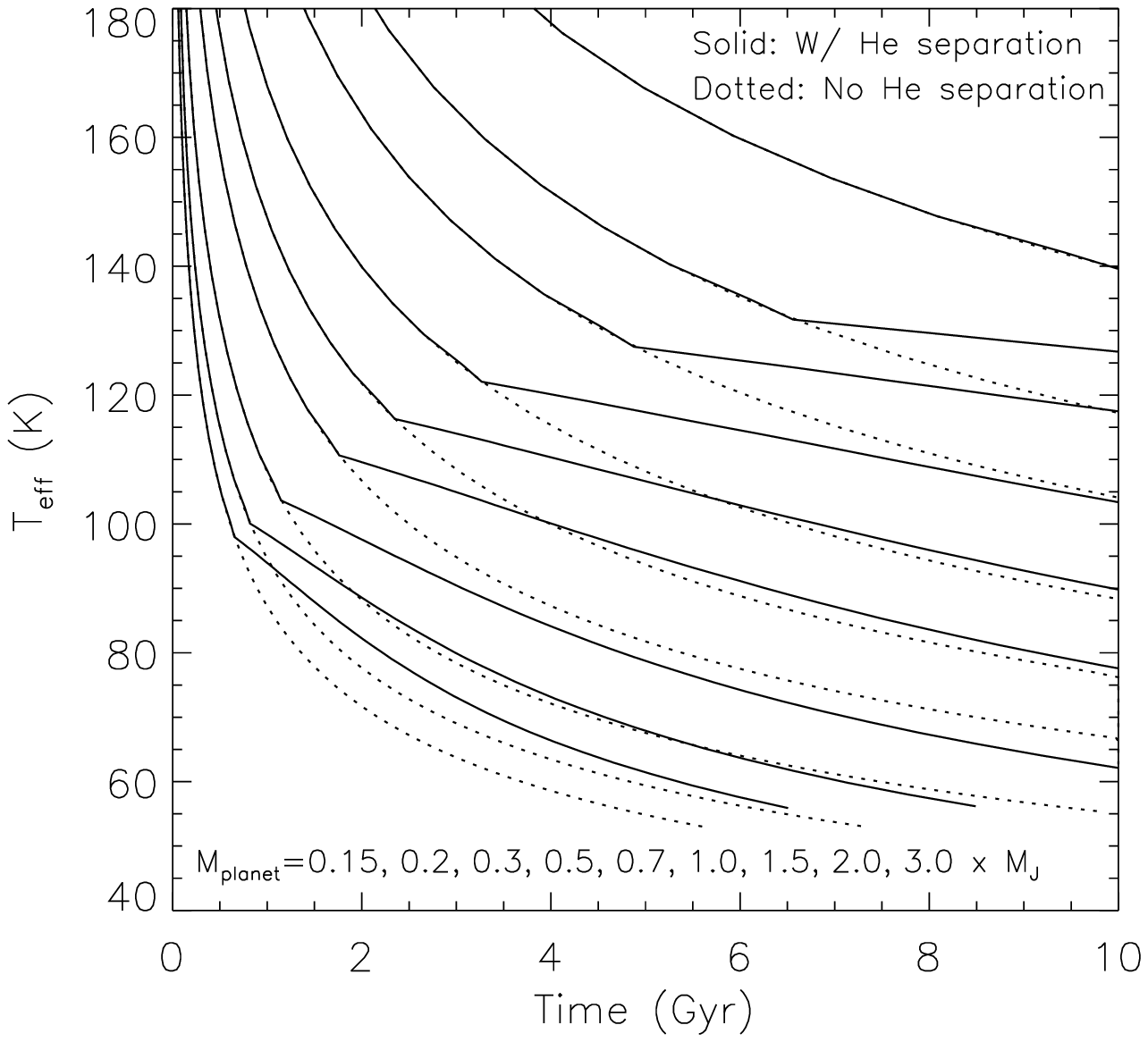}
\caption{Evolution of $T_{eff}$ for planets of mass 0.15 to 3.0 \mj\ for our standard models.  The top curve if for the highest mass planet while the bottom curve is for the lowest mass planet.
\label{figure:Teff}}
\end{figure}

\begin{figure}
\plotone{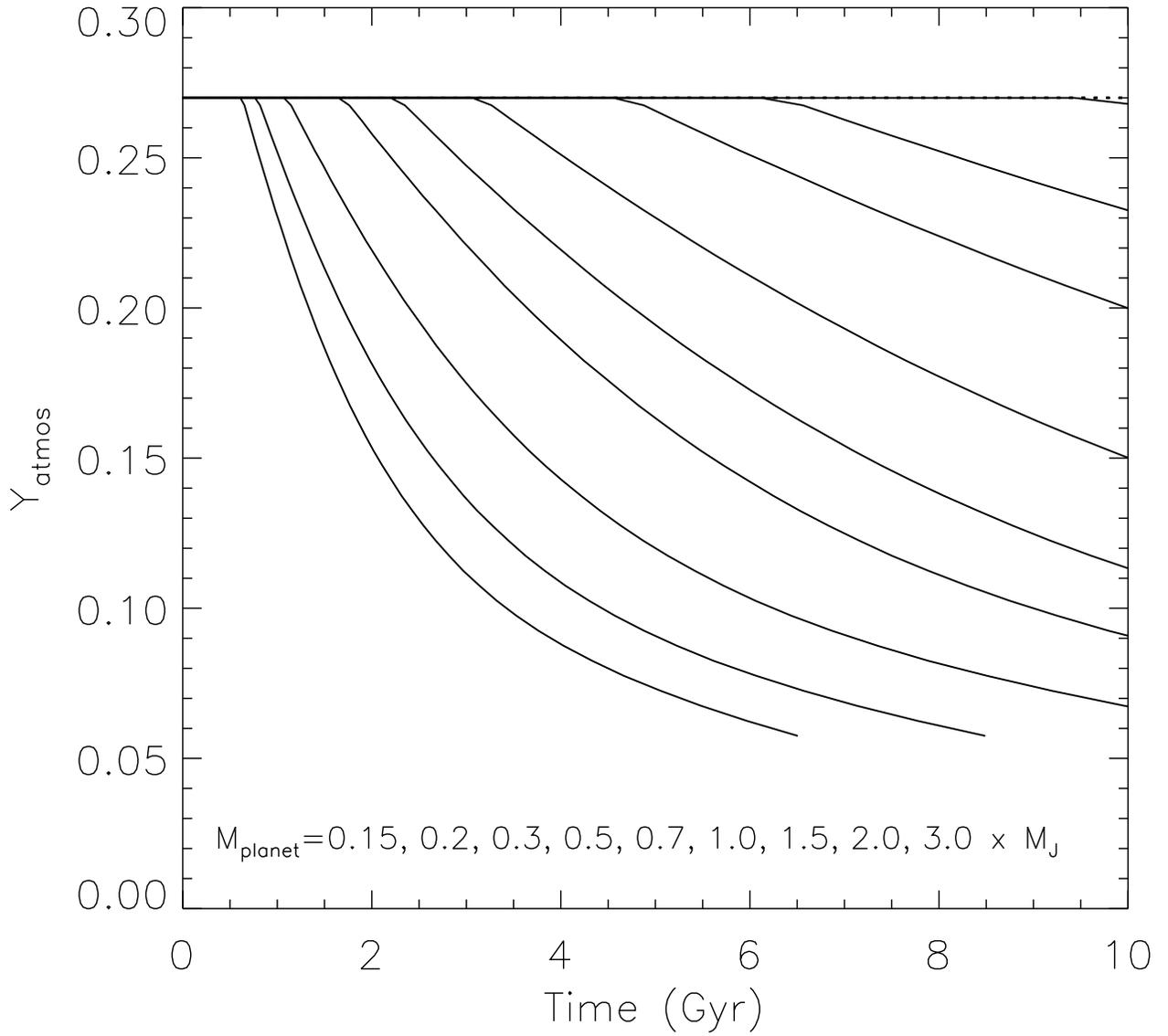}
\caption{Evolution of $Y_{atmos}$ (atmospheric helium mass fraction) for planets of mass 0.15 to 3.0 \mj\ for our standard models.  The lower the mass of the planet, the earlier the onset of helium phase separation, and the lower the final $Y_{atmos}$ at a given age.
\label{figure:Y}}
\end{figure}

\begin{figure}
\plotone{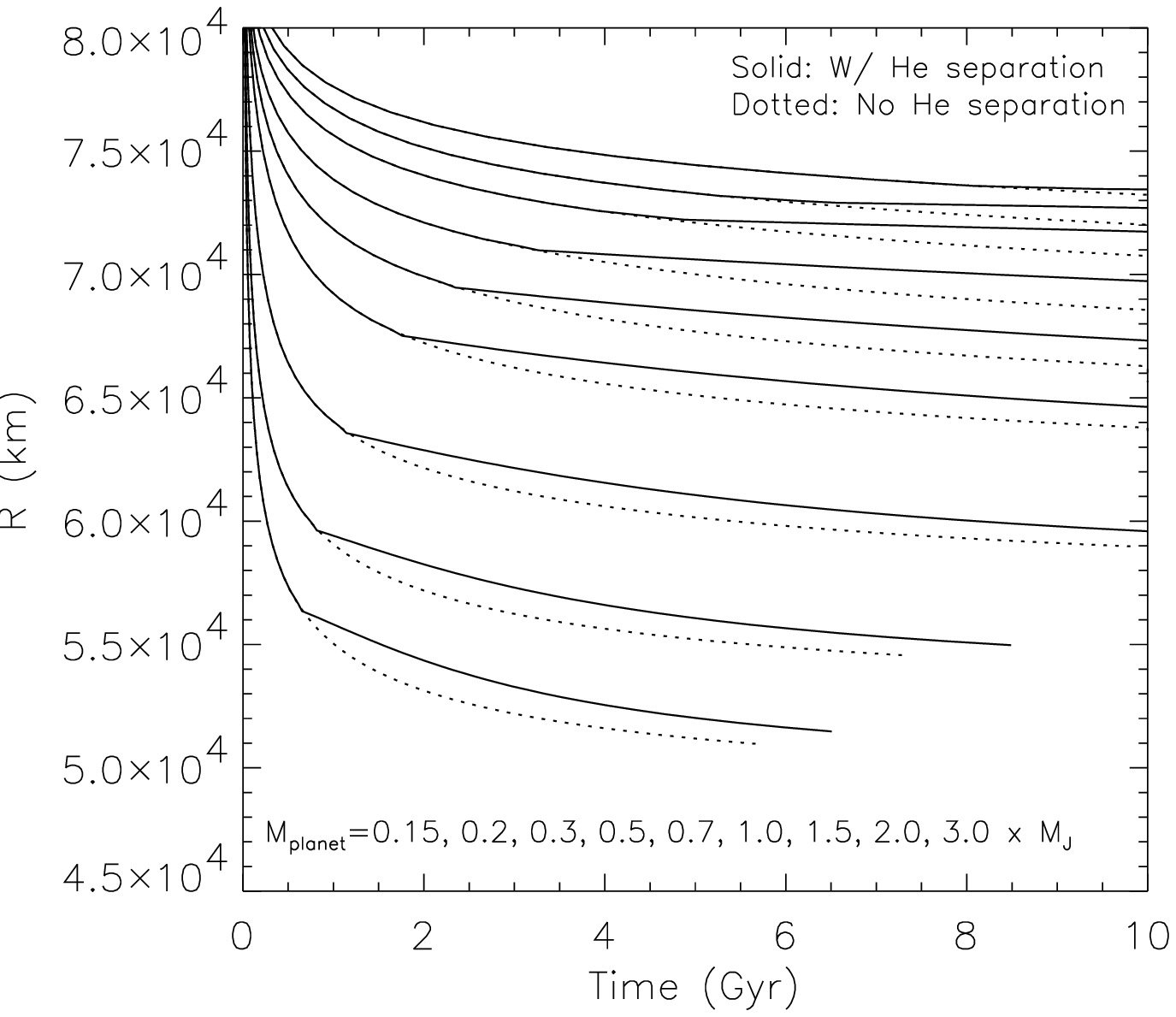}
\caption{Evolution of the radius of planets of mass 0.15 to 3.0 \mj\ for our standard models.  The top curve is for the highest mass planet while the bottom curve is for the lowest mass planet.
\label{figure:R}}
\end{figure}

The evolution of the low-mass EGPs we explore here can be drastically changed by the onset of phase separation of helium from hydrogen. \mbox{Figures~\ref{figure:L}}, \ref{figure:Teff}, \ref{figure:Y}, and \ref{figure:R} taken together map out our calculations for the luminosity, $T_{eff}$, $Y_{atmos}$, and radii of isolated EGPs.  Solid lines show inhomogeneous evolutionary models incorporating helium phase separation, while dotted lines are for homogeneous models.  A planet's luminosity falls at a much smaller rate, the gradual contraction of the planet is slowed, and the $Y_{atmos}$ falls as helium is lost from molecular and metallic regions, with the excess helium raining down to the planet's core.

The lowest mass planet we model is 0.15 \mj, which is 47.7 \me. This means $\sim$75\% of the planet's mass is hydrogen/helium envelope. This is approximately the lowest mass a planet can be while still having some liquid metallic hydrogen in its deep interior.  For comparison Uranus and Neptune are 14.5 and 17.1 \me, respectively.  This lower mass limit is dependent on the mass of the core and the exact pressure(s) at which hydrogen becomes metallic.  Later on we will show that coreless models for planets of this mass do not contain liquid metallic hydrogen.  For our standard 0.15 \mj\ model, the onset of helium separation is fairly early in the planet's evolution, at an age of 700 million years.  Of course our determination of this age for the onset of phase separation is dependent on the temperatures predicted by the \citet{SCVH} EOS and the planets' assumed initial helium mass fraction.

The evolution of planets of increasing masses proceed similarly.  Note that except for Figure~\ref{figure:L}, all figures show time linearly in Gyr.  This makes it clear how quickly giant planets fall from their initial large, highly luminous state.  At a given age, the higher the mass of the planet, the higher the entropy of the planet's interior.  Consequently, the greater the planet's mass, the later in the planet's evolution helium phase separation begins.  For our phase diagram, the liquid metallic region of the planet must reach a specific entropy of 6.11 $k_B$/baryon before helium begins to separate.

Figure~\ref{figure:L} shows the full evolution, starting from 1 Myr, of our range of planets.  Once helium phase separation begins, the luminosity of the lower mass planets (0.15 to 0.3 \mj) is increased by a maximum factor of 2.25 over the prediction of the homogeneous models.  For the higher mass planets this ratio is near 1.7.  The explanation for the effect being larger for the smaller mass objects can be understood coarsely in the following way.  For each planet, the same percentage of the planet's mass is falling onto the core.  A massive planet (say 2.0 \mj) that is 10 times more massive than a 0.2 \mj\ planet is only $\sim$~30\% larger in radius.  This means that the helium is falling a comparable distance, even though the masses differ by a factor of ten.  The planetary radius increases much more slowly with planet mass than the actual physical mass of helium raining down.  Consequently, the lower mass planets are affected to a larger degree.  The cooling rate in $T_{eff}$ per Gyr decreases by a factor of 4.5 upon the onset of phase separation.  A 1.0 \mj\ planet cools at a rate of -12 K Gyr$^{-1}$ just before the onset of phase separation and -2.6 K Gyr$^{-1}$ after.  This new cooling rate is maintained to within $\sim$~10\% past an age of 10 Gyr.

Changes in planetary $T_{eff}$ can be on the order of 10-15 K, when compared to homogeneous models, which is quite a marked difference, while radius increases are on the order of 1000-2000 km, which is somewhat small.  However, for the 1.5 and 2.0 \mj\ objects the onset of phase separation leads to a near halt of contraction of the planets.  Figure~\ref{figure:Y}, which shows $Y_{atmos}$ as a function of time gives one a feeling for how helium phase separation is proceeding in the planet as a function of time.  Unfortunately, $Y_{atmos}$ will never be an observable quantity except for in Jupiter and Saturn.  Notice that the 1.0 \mj\ model begins to lose helium to deeper layers at an age of 3.25 Gyr, 1.3 Gyr earlier than the age of the solar system.  The onset of phase separation will be delayed for all planets once the absorption of stellar photons is accounted for.

\subsection{Effects of Stellar Irradiation} \label{irrad}
The proceeding discussion is only correct for planets in isolation, such that no energy is absorbed from a parent star.  Very low mass planets in isolation may exist, but Gyr-old isolated planets seem unlikely targets for direct detection and characterization.  In our solar system, both Jupiter and Saturn reradiate more absorbed solar energy than internal energy.

\begin{figure}
\plotone{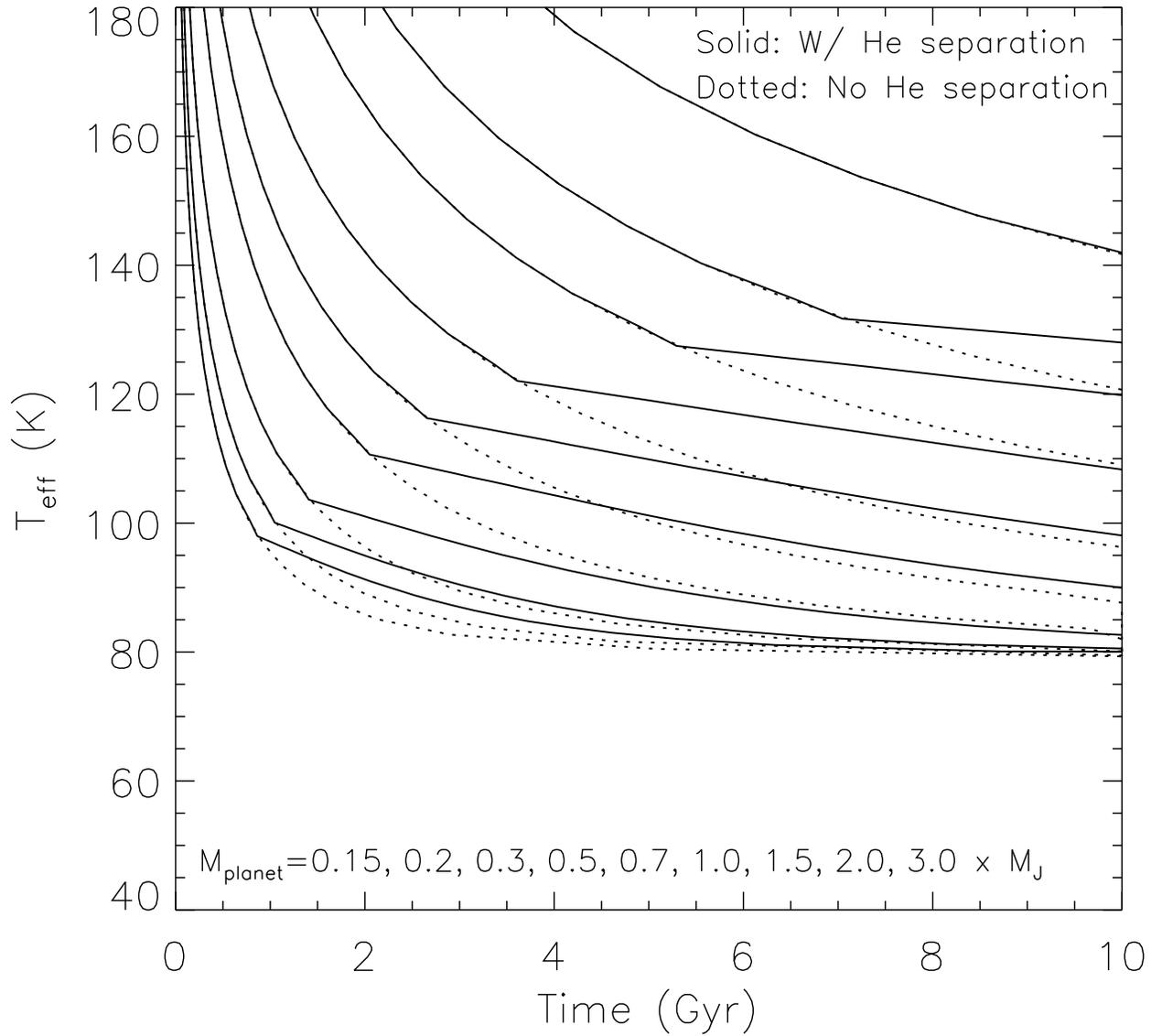}
\caption{Evolution of the $T_{eff}$ of irradiated model planets with a constant $T_{eq}$ of 79.2 K.  This corresponds to a distance of 10 AU from a constant luminosity 1.0 L$_{\odot}$ star.  Only the lowest mass planets reach their $T_{eq}$ in 10 Gyr.
\label{figure:T10}}
\end{figure}

\begin{figure}
\plotone{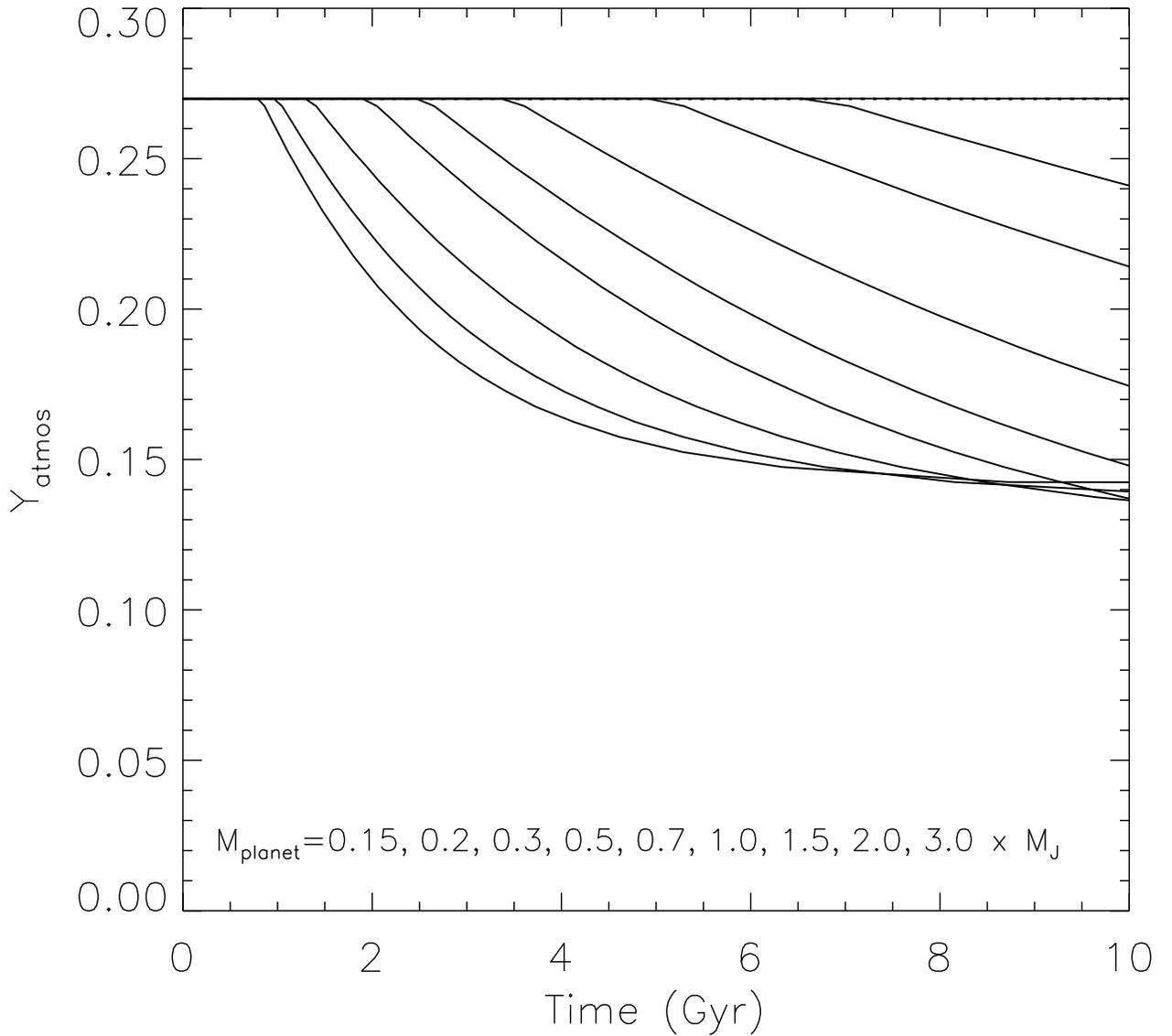}
\caption{The $Y_{atmos}$ for the model planets at 10 AU.  In our theory, when the lowest mass planets reach their $T_{eq}$, their evolution stalls, so no additional helium is lost to deeper layers.  The 0.15 and 0.2 \mj\ planets reach this point.  See the text for discussion on this issue.
\label{figure:Y10}}
\end{figure}

\begin{figure}
\plotone{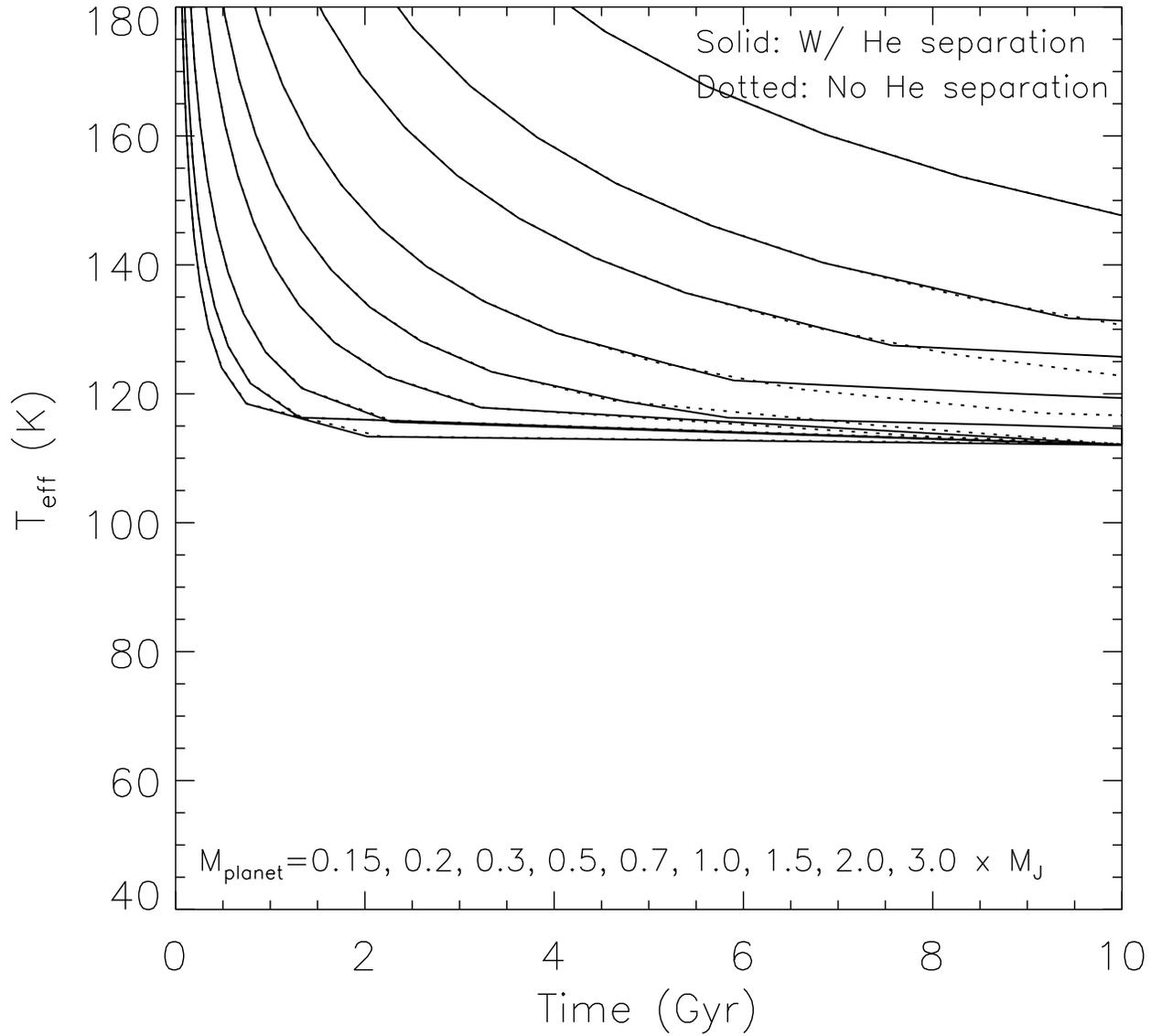}
\caption{Evolution of the $T_{eff}$ of irradiated model planets with a constant $T_{eq}$ of 112.0 K.  This corresponds to a distance of 5 AU from a constant luminosity 1.0 L$_{\odot}$ star.  The evolution of all planets is significantly affected, with the lowest mass planets reaching $T_{eq}$ in less than a few Gyr.  The 0.15 to 0.5 \mj\ planets reach their $T_{eq}$ before the onset of helium phase separation.
\label{figure:T5}}
\end{figure}

\begin{figure}
\plotone{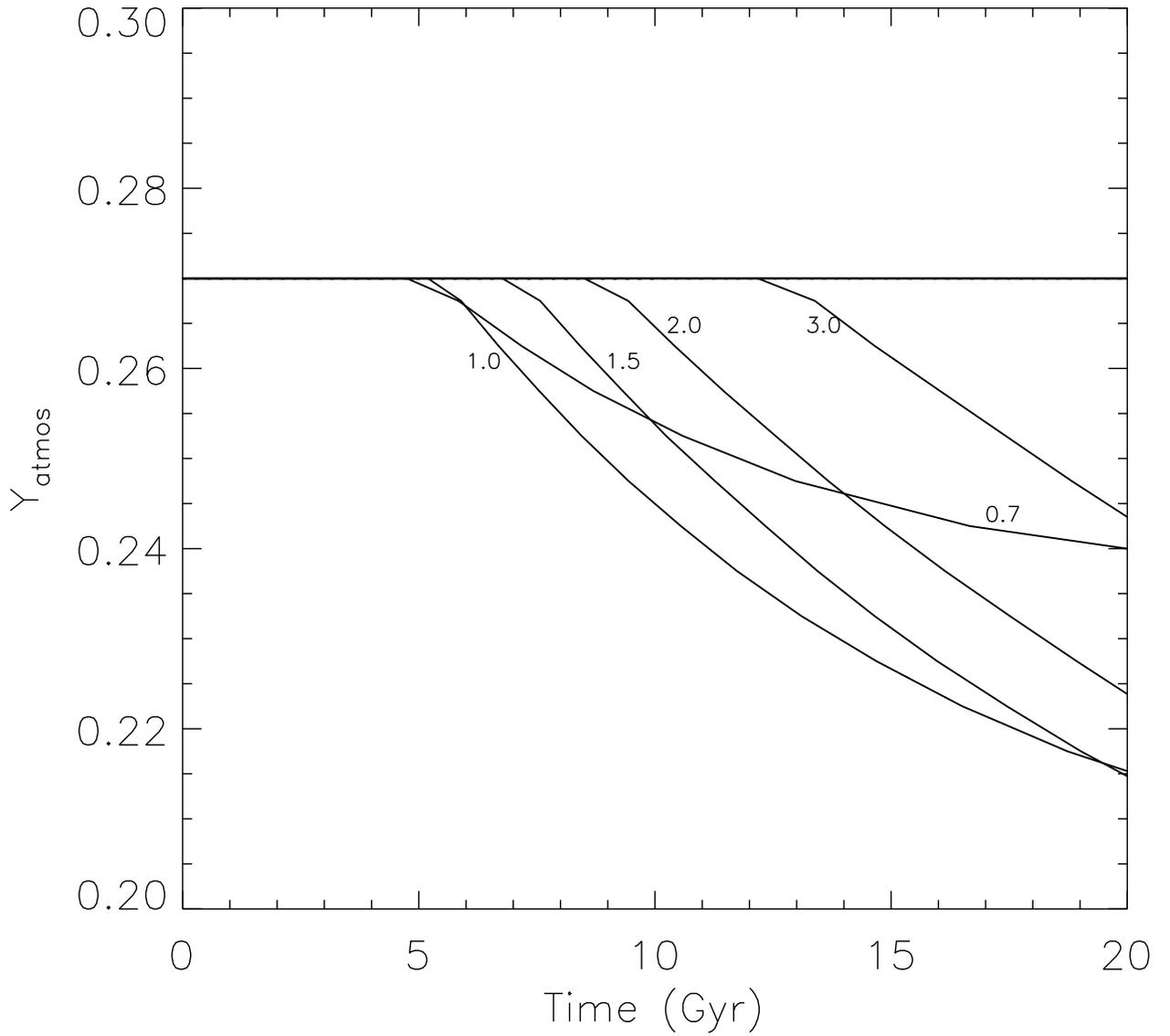}
\caption{The $Y_{atmos}$ for the model planets at 5 AU, up to an age of 20, rather than 10, Gyr.  Curve labels are the planets' masses in \mj.  Due to the stellar heating 0.15 to 0.5 \mj\ planets reach their $T_{eq}$ before the onset of phase separation, while the 3.0 \mj\ planet stays too warm for 12 Gyr.  Once a planet reaches its $T_{eq}$, its thermal radiation is dominated by thermalized stellar photons, rather than intrinsic energy.  This is most obvious for the 0.7 \mj\ planet as its $Y_{atmos}$ nearly reaches an asymptotic value just below 0.24.
\label{figure:Y5}}
\end{figure}

In order to investigate how stellar irradiation may alter the effects of helium phase separation, we place our same model planets at a distance of 10 and 5 AU from a star with a constant luminosity of 1.0 L$_{\odot}$.  We make use of theory of \citet{Hubbard77} outlined in Section \ref{models}.  The accuracy of this method breaks down when a planet's $T_{eff}$ reaches its $T_{eq}$, however.  As has been discussed in relation to the intensely irradiated EGPs such as HD 209458b \citep{Guillot02}, the \citet{Hubbard77} theory, when making use of a model atmosphere grid for isolated planets, predicts a halting of the contraction of a planet at $T_{eff}=T_{eq}$ (and hence a halting of the cooling of their interiors) but in reality a planet likely continues to contract while maintaining a constant $T_{eff}=T_{eq}$, as the radiative region of the planet's atmosphere expands to encompass higher pressures.  This should be kept in mind when interpreting Figures \ref{figure:Y10} and \ref{figure:Y5}, which show a likely unphysical halting of phase separation.  The calculations of $T_{eff}$ in Figures \ref{figure:T10} and \ref{figure:T5} are accurate, as $T_{eff}$ will eventually fall to the $T_{eq}$ value.

Currently, Jupiter and Saturn have essentially the same bond albedo of 0.343.  To date, evolutionary models of these planets have used this value for their entire evolution, which is overly simplified.  Since we are not calculating consistent atmosphere models for various stellar distances and planet gravities, we too assume our model planets have a bond albedo of 0.343.  Although progress in modeling giant planet albedos has been made \citep{Sudar00}, great uncertainties in the modeling of equilibrium condensate clouds in EGP atmospheres remain \citep{AM01,Cooper03}.  Since an understanding of \emph{nonequilibrium} condensates is needed to account for the albedos of Jupiter and Saturn, much work still needs to be done.  Prospects for an accurate understanding of EGP albedos in general will be helped by detections of reflected light from EGPs.

Since the store of internal energy for the lowest mass planets is the smallest, and they consequently have the lowest $T_{eff}$s at a given age, they will be the most affected by stellar irradiation.  At a distance of 10 AU, a planet's $T_{eq}$ is 79.2 K and at 5 AU it is 112.0 K.  For reference, if one were instead to consider a M5V star of luminosity L=0.22 L$_{\odot}$, these $T_{eq}$s would correspond to distances of 2.3 and 4.6 AU.

Figures~\ref{figure:T10} and \ref{figure:Y10} show how the planets' $T_{eff}$ and $Y_{atmos}$ are affected by stellar irradiation at 10 AU.  The onset of helium phase separation is delayed by $\sim$~300 Myr, but the planetary evolution proceeds in a fashion similar to the isolated case for the more massive planets.  However, the less massive planets are qualitatively affected by the stellar heating.  The 0.15 and 0.2 \mj\ planets reach their $T_{eq}$ in 7 and 8 Gyr, respectively, and at that point their energy budget is dominated by the stellar, rather than intrinsic, source.  The age of the 1.0 \mj\ planet at the onset of helium phase separation is 3.5 Gyr, still only $\sim$~77\% of the age of the solar system.

The effects of stellar irradiation are even more clear in Figures~\ref{figure:T5} and \ref{figure:Y5}.  These figures show planets with $T_{eq}$=112.0 K.  The planets with masses less than 0.7 \mj\ reach their $T_{eq}$ before helium phase separation can even begin.  With their evolution stalled at this early age, our theory predicts that they never cool enough for helium to become immiscible in their interiors.  However, as these planets likely do continue to contract at $T_{eff}=T_{eq}$, a more correct statement would be that they begin phase separation well after their $T_{eff}$ is dominated be absorbed stellar energy rather than internal energy.  The time of onset of phase separation is extended about 2 Gyr for all the planets.  The 0.7 \mj\ planet has just begun to lose helium to deeper layers when its evolution is stalled, as is seen in Figure~\ref{figure:Y5}.  For the 1.0 \mj\ object, helium phase separation begins at an age of $\sim$~5.4 Gyr.

For both sets of irradiated models, the planetary radii can be calculated by reading off the mass and $T_{eff}$ from Figure~\ref{figure:T10} or \ref{figure:T5}, consulting Figure~\ref{figure:Teff} for the age of an \emph{isolated} model for that $M$/$T_{eff}$, and then reading the radius off Figure~\ref{figure:R} at the isolated planet age.  This works because of our approximation that the stellar flux is absorbed in the convective region of the planet.

\subsection{Effects of Alternate Core Masses} \label{cores}
The current mass of Jupiter's and Saturn's heavy element cores can be constrained with static models that must match each planet's mass, rotation rate, radius at 1 bar, temperature at 1 bar, and gravitational moments J$_2$, J$_4$, and J$_6$, using given EOSs for hydrogen, helium, and heavier elements.  The most recent estimates for these core masses, taking into account uncertainties in all parameters, give a core mass of 0-10 \me\ for Jupiter and 10-20 \me\ for Saturn \citep{Saumon03}.

Our understanding of Jupiter and Saturn led us to choose 10 \me\ as a realistic core mass for our standard models.  Prospects for determining whether or not EGPs possess cores are uncertain.  For transiting EGPs, if a detection of a planet's oblateness could be measured from a light curve, and its rotation rate derived with some other method, some indication of the amount of central concentration of the mass could be estimated \citep{Barnes03, Seager02}.  A more promising route may come from obtaining the radii of low-mass transiting giant planets, where differences in radii for planets with and without cores will be large \citep{Bodenheimer03}.  However, a number of unknowns may complicate this picture, as uncertainties in stellar age, radius, and mass translate directly into uncertainties in planetary age, radius, and mass.  These effects may be somewhat difficult to untangle.

\begin{figure}
\plotone{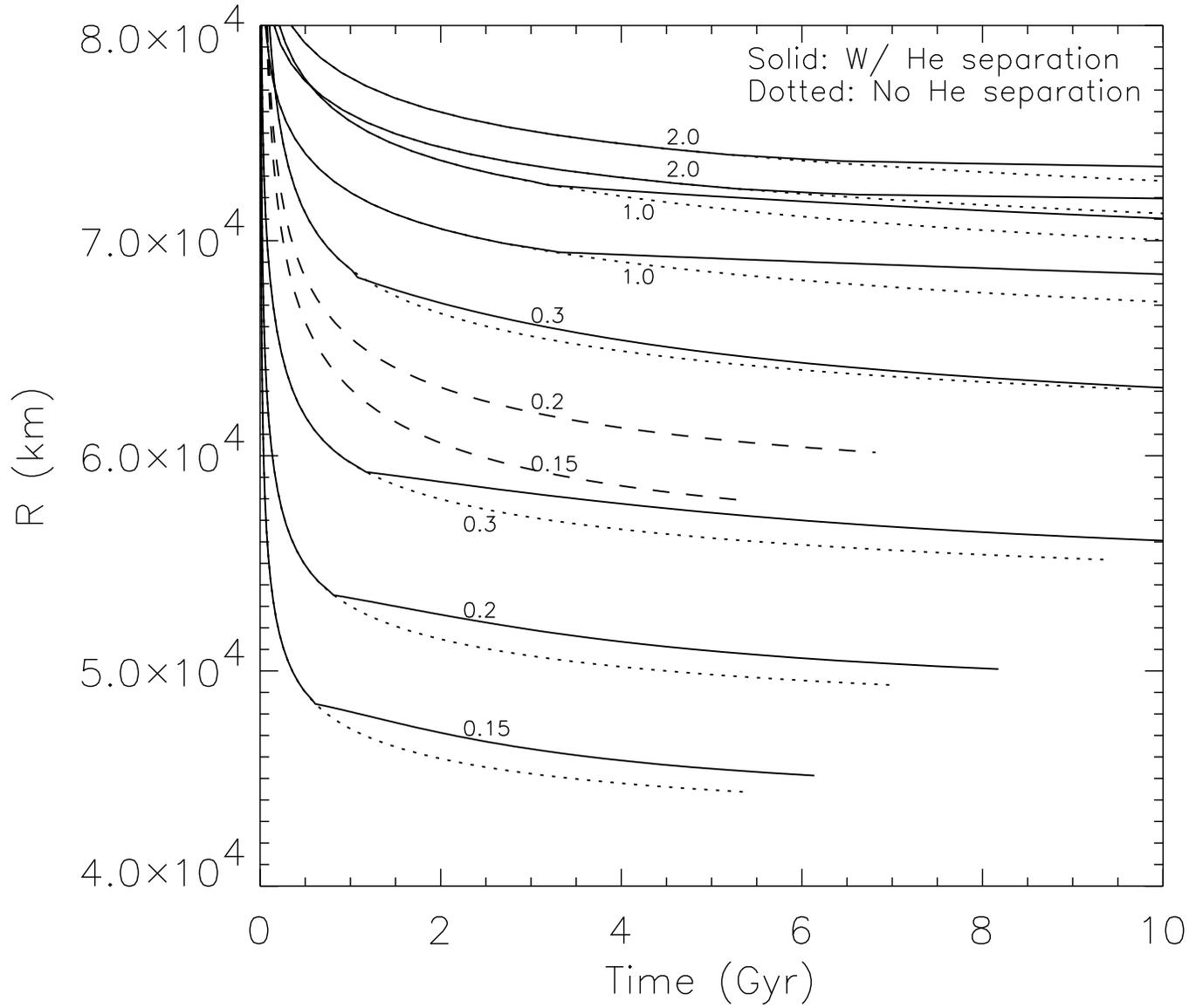}
\caption{The evolution of isolated planets' radii as a function of time for coreless models and models with heavy element cores of 20 \me.  Planets are labeled in \mj: 0.15, 0.2, 0.3, 1.0, and 2.0.  For each mass pair, the planet with the larger radius is the coreless model.  The coreless model planets of mass 0.15 and 0.2 \mj\ (dashed lines) do not have central pressures high enough for liquid metallic hydrogen to form ($\sim$~2 Mbar) at any age, and therefore helium will not become immiscible. 
\label{figure:Rcores}}
\end{figure}

Therefore, it is worthwhile to investigate what effects larger or smaller core masses may have on the evolution of EGPs.  Our alternate models have core masses of zero and 20 \me.  As one would presume, and can be seen in \mbox{Figure~\ref{figure:Rcores}}, the most obvious effect is on the radii of the planets.  A larger mass of heavy elements will significantly reduce a planet's radius.  This difference can be 10,000 km for the lower mass planets.  (This is true whether the heavy elements are in the core or are uniformly mixed.)  As the model planets increase in mass, the difference between zero and 20 \me\ of heavy elements becomes a decreasing smaller percentage of the planets' mass, and therefore has a corresponding smaller effect on the planetary radii.

As was noted in \mbox{Figure~\ref{figure:phased}}, EGPs below a certain mass will not have a high enough central pressure to possess any liquid metallic hydrogen.  Therefore, these planets cannot undergo phase separation of helium at these temperatures, because no liquid metallic hydrogen exists for the helium to become immiscible in.  (Helium likely becomes immiscible in \emph{molecular} hydrogen at some temperature, but this is almost assuredly at temperatures much lower than those found in giant planets \citep{SS77a}, except perhaps near $\sim$~1.0 Mbar.)  At a given planet mass, the greater the mass of the core the larger the central pressure in the core $and$ the larger the pressure at the core/envelope boundary.  Since the pressure at which hydrogen turns metallic is currently unknown we simply choose 2 Mbar as the transition pressure, independent of temperature.  Our coreless 0.15 and 0.2 \mj\ planets do not reach this central pressure in 10 Gyr, while the coreless 0.3 \mj\ planet does in $\sim$~1 Gyr, so helium immiscibility proceeds as described earlier.  (Current evidence indicates the hydrogen insulator/metallic transition is likely continuous, rather than 1st order, and theory predicts this transition is also a function of temperature.  See \mbox{Figure~\ref{figure:phased}} and references in the caption.)

\begin{figure}
\plotone{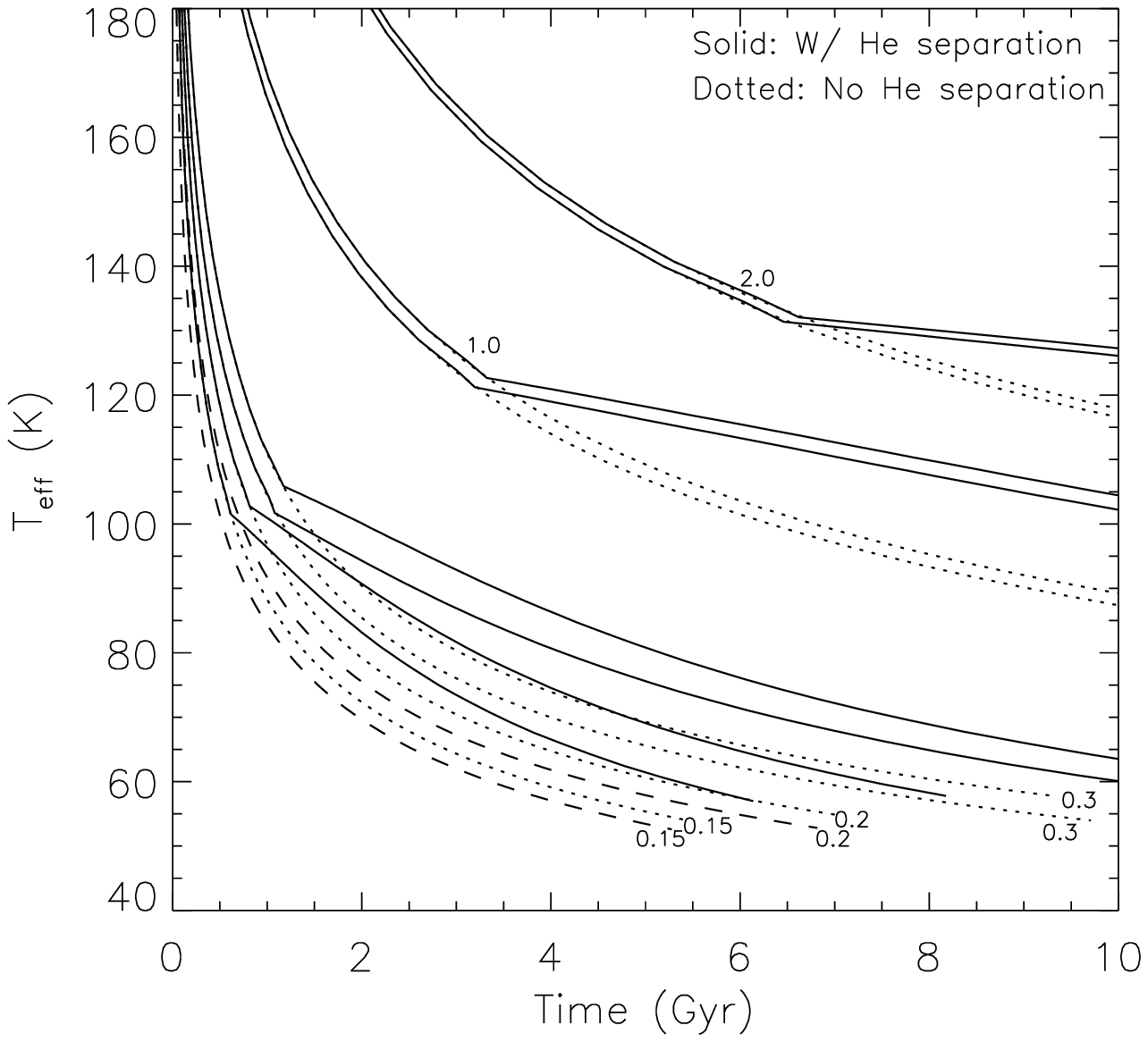}
\caption{The evolution of isolated planets' effective temperatures as a function of time for coreless models and models with heavy element cores of 20 \me.  Planets are labeled in \mj: 0.15, 0.2, 0.3, 1.0, and 2.0.  For each mass pair, the planet with the lower $T_{eff}$ is the coreless model.  The coreless model planets of mass 0.15 and 0.2 \mj\ (dashed lines) do not have central pressures high enough for liquid metallic hydrogen to form ($\sim$~2 Mbar) at any age, and therefore helium will not become immiscible.  
\label{figure:Tcores}}
\end{figure}

This effect leads to an interesting bifurcation in the evolution of the lowest mass EGPs.  The 0.15 and 0.2 \mj\ coreless planets possess no liquid metallic hydrogen and cool homogeneously during their entire evolution, unaffected by helium phase separation.  The models with 10 and 20 \me\ cores are affected by helium phase separation, and consequently have quite different $T_{eff}$s after several Gyr.  This effect can be seen in \mbox{Figure~\ref{figure:Tcores}} which shows the evolution of coreless EGPs and those with 20 \me\ cores.  The models with large cores can have $T_{eff}$s $\sim$~15 K higher than coreless models---a difference of $\sim$~20\% due to core size alone!  The coreless 0.15 and 0.2 \mj\ planets are shown as dashed lines for clarity.  It should be noted that if helium immiscibility does occur at planetary temperatures in very dense conducting molecular hydrogen near $\sim$~0.7-1 Mbar, this bifurcation in evolution would occur at slightly lower masses.

\section{Discussion} \label{discuss}
\subsection{Ammonia Cloud Formation Timescales}
From the proceeding discussion and figures it is clear that helium phase separation can be a substantial additional energy source in EGPs, therefore making these objects more luminous than one would predict from homogeneous models.  This increased luminosity will also delay the time until the formation of condensates in the planets' atmospheres.  For our phase diagram considered here, since helium phase separation does not begin until the EGPs are rather cold, most abundant condensates will have formed and moved to high pressures (well below the visible atmosphere) before helium becomes immiscible in a planet's interior.  However, the formation of ammonia clouds is during or after the onset of helium phase separation, and consequently, timescales associated with these clouds will be delayed due to this additional energy source. This will change the time these planets are Class I EGPs, as defined by \citet{Sudar00} and \citet{Sudar03}. 

We calculated the time necessary for our planetary adiabats, at a pressure of 1 bar, to reach the condensation curve of ammonia, as taken from \citet{AM01}.  This is the time necessary for the ammonia cloud base to reach 1 bar.  We were interested in whether the additional energy source due to helium phase separation could delay or stall the formation of clouds in a planet's atmosphere.  This calculation is only correct if the planets are adiabatic by a pressure of 1 bar in the atmosphere.  This seems to hold true for Jupiter and most EGP atmosphere models for isolated or moderately irradiated planets  \citep{Sudar03}.  Since we do not compute atmosphere models, we cannot describe the effects on cloud formation at higher altitude, lower pressure regions of the atmosphere that do not lie on the interior adiabat.  As the pressure decreases, atmospheric \emph{T-P} profiles become more isothermal.  But here we outline the general effect.

\begin{figure}
\plotone{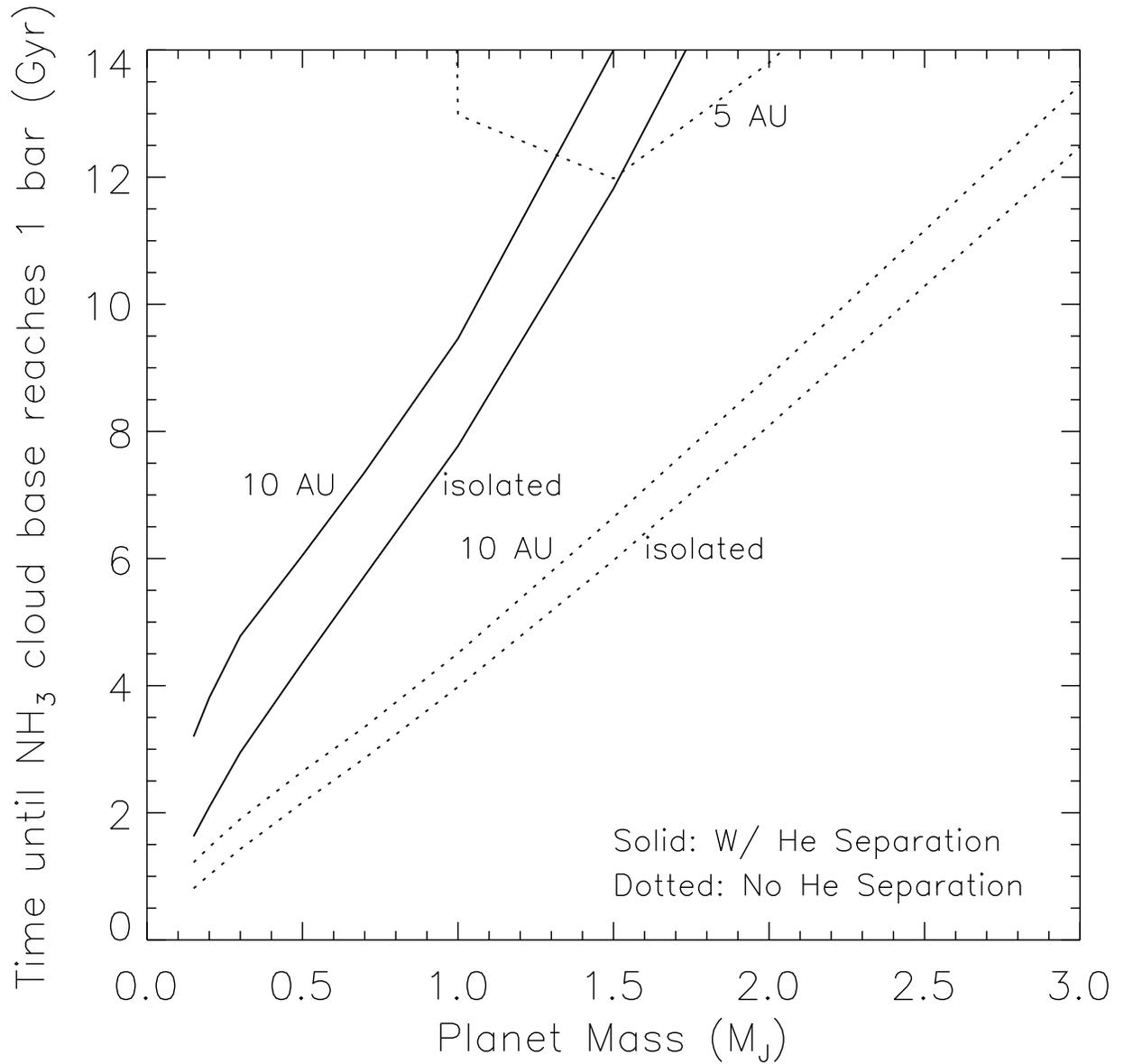}
\caption{Helium phase separation occurs before or (more likely) during the formation of ammonia clouds, which will delay their settling to higher pressures in the planets' atmospheres as they cool.  This figure shows, as a function of planet mass and irradiation, how long it takes ammonia clouds to reach 1 bar pressure in the planets' atmospheres.  If the effects of phase separation are included, planets at 5 AU will never form ammonia clouds that reach 1 bar.
\label{figure:Tcloud}}
\end{figure}

Figure~\ref{figure:Tcloud} shows, for our isolated and irradiated models, how phase separation and stellar irradiation change the time for the ammonia clouds to reach 1 bar.  The delay due to phase separation is always greater than $\sim$~1 Gyr, and can be 6 Gyr or longer for more massive planets.  Planets within 5 AU of the parent star only marginally reach this point with no helium phase separation.  The lowest mass planets reach their $T_{eq}$ before the ammonia clouds can reach 1 bar.  When phase separation is included this is true for all planets, independent of mass.  What remains unclear without atmosphere models is whether helium phase separation delays the start of the formation of the ammonia clouds, or alternatively, if the clouds have already begun forming before phase separation begins, and the helium phase separation merely greatly extends the timescale the clouds are in the visible atmosphere.  If we use Saturn as a guide, which has visible clouds that are thought to be composed of condensed ammonia, it is likely the latter, although the position of the condensation curve of ammonia will depend on the planet's nitrogen abundance.  It seems that due to the predicted delay in cooling, ammonia clouds will reside in a planet's visible atmosphere for several additional Gyr.

\subsection{Applications to the Cooling of Jupiter and Saturn}
As can be seen from the models in Section \ref{results}, the amount of energy absorbed by a giant planet due to stellar irradiation has a critical effect on the time for the onset of helium phase separation.  The model planets at 5 AU reach helium phase separation 2 Gyr later than the isolated planets.  The fact that we highlighted the evolution of 1.0 \mj\ planets especially is no accident.  Jupiter's $Y_{atmos}$ is only 86\% of the $Y_{protosolar}$, and this number, along with the $Y_{atmos}$ of Saturn, must be explained in any consistent evolutionary history of the planets.

\citet{Hubbard99} showed that the onset of helium phase separation in Jupiter leads to model planet ages greater than the age of the solar system when Jupiter reaches $T_{eff}=124.4$ K.  However, \citet{Hubbard99} (and \pf) made the same assumptions outlined in Section \ref{models}.  They assume all photons that are absorbed are done so in adiabatic layers, and that Jupiter's atmospheric $T-P$ profile does not deviate from that of an isolated object.  From the Galileo entry probe we know that this is not the case.  In addition, these papers assume Jupiter's Bond albedo is a constant 0.343 with time.  However, if Jupiter's Bond albedo was larger in the past, when H$_2$O clouds were higher in the planet's atmosphere, its bond albedo could be somewhat larger and less solar radiation would be absorbed.  This would lead to faster cooling and could admit into Jupiter models some phase separation before an age of 4.6 Gyr.  While evolutionary models consistently incorporating incident stellar radiation have been calculated for HD 209458b \citep{Baraffe03,Burrows03}, similar work has not yet been done for Jupiter and Saturn.  

\subsection{Future Work and Observations}
Since we have yet to work out a consistent evolutionary history for Jupiter and Saturn that explains the atmospheric abundance of helium in both of these planets, the possibility that a different element could be undergoing phase separation must be entertained.  This could be happening in addition to, or instead of, helium's phase separation.  Possibilities include neon, which was found in only 1/10th the solar abundance in Jupiter's atmosphere (no value is known for Saturn) and oxygen, since a recent model of Jupiter's formation \citep{Gautier01a, Gautier01b} predicts H$_2$O may be \emph{at least} 9.4 times more abundant in Jupiter than it is in the Sun.  This fairly large abundance could lead to immiscibility temperatures of several thousand K at megabar pressures, but no detailed oxygen/hydrogen phase diagrams have yet been calculated at these pressures.

If a radial velocity candidate planet is directly imaged in the coming years, without an accurate mass determination by another method, it will be important to use inhomogeneous evolutionary models that include helium phase separation when trying to deduce the planet's mass from model evolution tracks.  In addition, if astrometry missions such as SIM lead to EGP detections (with mass determinations) around stars, it will be worthwhile to include the effects of phase separation when trying to understand the sensitivity needed to image these planets.  A measured luminosity of a low-mass EGP with a known mass would be an important observational test of our suggested phase diagram.  In addition, these larger intrinsic luminosities should be taken into account when calculating accurate atmosphere models for EGPs, and their corresponding emission and reflection spectra.

While younger planets have the advantage of being several orders of magnitude brighter if their systems are only millions of years old, planets found by radial velocity or astrometry perturbations will be attractive targets because planetary orbital parameters will already be known.  These planets that orbit at several AU and beyond around their parent stars, if one can use Saturn and Jupiter as a guide, may well be $\sim$~twice as luminous as current homogeneous models predict.  Of the currently available planetary candidates, $\epsilon$ Eri b and 55 Cnc d at first glance seem to have the greatest likelihood of having helium phase separation currently affecting their evolution.  However, even though $\epsilon$ Eri b has a low minimum mass (0.92 \mj\ \citep{Hatzes00}), its age is likely only 1 Gyr \citep{Drake93}.  55 Cnc b, although it orbits with a semimajor axis of 5.9 AU, has a minimum mass of 4.05 \mj\ \citep{Marcy02}, making it too massive to undergo helium phase separation.

The calculations presented here are an early step in exploring phase separation in EGPs.  No experimental data are yet available for high pressure hydrogen-helium mixtures at planetary temperatures.  Advances in understanding phase separation will occur through high pressure experiments, theory, further detailed evolutionary models of Jupiter and Saturn, and an accurate determination of Saturn's atmospheric helium abundance.  While we believe our calculations for planets less massive than Jupiter should be reasonably accurate, since we have calibrated the theory to these planets, the extrapolation up to 3.0 \mj\ is uncertain, since it is not clear that the immiscibility curves we show in Figure \ref{figure:phased} maintain this positive slope at pressures greater than tens of Mbar.  A turnover in the immiscibility curves would lead to helium redissolving in deeper liquid metallic hydrogen layers, rather than settling onto the core.

\section{Conclusions}
Using the phase diagram described in \pf, which is calibrated to Jupiter and Saturn and allows both planets to reach their known ages and effective temperatures, we have explored the effects that helium phase separation will have on a variety of EGPs.  The additional energy liberated as helium rains to deeper layers of a planet will significantly delay the cooling and contraction of giant planets.  Once helium phase separation is underway, our inhomogeneous evolutionary models predict luminosities $\sim$~2 times greater than predictions from homogeneous models.  This will make these giant planets in the 0.15 to 3.0 \mj\ mass range somewhat easier to detect that has been previously thought.  Improvements in understanding the evolution of these EGPs will come through a better understanding of our closest giant planets, Jupiter and Saturn.

\acknowledgments
Electronic files of the evolutionary models presented here can be obtained by contacting JJF.  We thank Adam Burrows, David Sudarsky, Jonathan Lunine, Ivan Hubeny, and Jason Barnes for many interesting conversations while this work was underway.  JJF is funded by a NASA GSRP grant and WBH by NASA PG\&G grant NAG5-13775.

\newpage

\newpage

\newpage

\newpage

\newpage

\newpage 

\newpage

\newpage

\newpage

\newpage

\newpage

\newpage

\newpage

\newpage

\newpage

\end{document}